\documentclass[aps,showpacs,nofootinbib]{revtex4}
\usepackage{epsf,latexsym}
\usepackage{amsfonts}

\begin{document}

\title{Quantum metric for null separated events and spacetime atoms}
\author{Alessandro Pesci\footnotetext{e-mail: pesci@bo.infn.it}}
\affiliation
{INFN Bologna, Via Irnerio 46, I-40126 Bologna, Italy}

\begin{abstract}
Recently, a proposal has been made to figure out the
expected discrete nature of spacetime at the smallest scales
in terms of atoms of spacetime,
capturing their effects
through a scalar $\rho$,
related to
their density,
function of the point $P$
and vector $v^a$ at $P$.
This has been done
in the Euclideanized space
one obtains through analytic continuation
from Lorentzian sector at $P$.
$\rho$ has been defined in terms of a peculiar `effective' metric $q_{ab}$,
of quantum origin,
introduced for spacelike/timelike separated events.
This metric stems from requiring that $q_{ab}$ coincides
with $g_{ab}$ at large (space/time) distances, but gives finite distance
in the coincidence limit,
and implements directly this way one single, very basic 
aspect associated to any quantum description of spacetime:
length quantization.
Since the latter appears
a quite common feature in the
available quantum descriptions of gravity,
this quantum metric $q_{ab}$
can be suspected to have a rather general scope
and to be re-derivable (and cross-checkable) in various specific quantum
models of gravity, even markedly different one from the other.

This work reports on
an attempt to introduce a definition
of $\rho$ not through the Euclidean
but directly in the Lorentz sector.
This turns out to be not a so trivial task,
essentially because of the null case,
meaning when $v^a$ is null,
as in this case it seems we lack even a concept of $q_{ab}$.
A notion for the quantum metric $q_{ab}$
for null separated events is then proposed
and an expression for it is derived.
From it,
a formula for $\rho$ is deduced,
which turns out to coincide with what obtained
through analytic continuation.
This
virtually completes
the task of having quantum expressions
of any kind of spacetime intervals,
with, moreover, $\rho$ defined directly
in terms of them (not in the Euclideanized space).
\end{abstract}


\maketitle

$ $
\section{Staying in the Lorentz sector}

In the context of the attempts to provide a quantum theory of gravity
or to describe spacetime quantum-mechanically,
some works \cite{KotE, KotF, StaA}
have lately proved it quite useful to introduce 
a peculiar sort of effective or quantum metric $q_{ab}$,
also called qmetric,
which acts to some extent as a metric at the same time allowing for
the existence of a finite limiting distance $L$
between two events in their coincidence limit.
It implements this way intrinsic discreteness of spacetime,
still not abandoning the benefits, for calculus, associated
to a continuous description of spacetime.
One point of merit of this qmetric approach appears
to be its genericity.
Indeed, the quantum description it offers,
does not come from a specific quantum theory of gravity
but arises instead straight from
simply requiring length quantization,
a feature, this, one is likely to find
in most specific models,
and which has as such the status of quite a generic expectation when
quantizing gravity.
In Loop Quantum Gravity (LQG) \cite{AshD, ThiA, RovQ} for example,
quantization
goes through the discretization of the classical theory
(general relativity) and the introduction of a quantum theory
associated to this discretization.
We do get length quantization in it; this however not directly,
but as a consequence of the general quantization procedure just mentioned.
What we can say is that,
concerning length quantization effects,
it seems in principle we can
compare what predicted by LQG and by the qmetric,
with the predictions of the latter coming from length quantization
without any specific theory associated with,
and those of the former
coming instead
from the quantum framework provided by the specific theory.
This means that the results one can obtain
with the qmetric approach,
could have wide range applicability within
the various specific quantum gravity models,
no matter how much they may differ one from the other
in their starting assumptions and perspective
(for example, whether the quantum theory of gravity
has to come from the quantization of the classical theory of gravity
or hinges instead on some, as yet unknown/untested, physics
at Planck scale)
and should
be in principle recoverable in any one of them
(and results in this sense have been
reported in \cite{CouA, CouB}).

One result one gets thanks to the quantum metric $q_{ab}$ is
the possibility to provide a notion 
of degrees of freedom or of number of (quantum) states
of spacetime at a point \cite{Pad04, Pad06, Pad08, Pad10},
fact which paves the way to a statistical description
of field equations, and then to express the basic tenets of gravity
using as proper language thermodynamics (as opposed to geometry)
\cite{Pad04}.
This endows a previous statistical derivation of field equations
\cite{PadG, PadF}
and the notion of horizon microscopic degrees of freedom
\cite{PadK, PadL}
as well as recent results 
connecting
so-called black hole chemistry
\cite{ManB, ManA}
with horizon degrees of freedom
\cite{Var},
all arising from macroscopic spacetime thermodynamics,   
with a microscopic mechanism seemingly able 
to directly justify these degrees of freedom.
Key to the notion of degrees of freedom or of number of states
of quantum spacetime is a quantity, denoted here $\rho$,
defined in terms of $(D-1)$-dimensional areas
(spacetime is assumed $D$-dimensional) of hypersurfaces
formed by points at assigned
distance from some point $P$
in the space coming from
Euclideanisation of original spacetime around $P$.
The basic feature about it
is that, according to the effective metric,
these $(D-1)$-areas remain finite
in the coincidence limit of the hypersurfaces shrinking to $P$
\cite{Pad04} (and clearly, one would expect some analogous results
do hold true in Lorentzian sector). 

This of Euclideanisation
might be a point of merit,
providing insight perhaps into what the structure of the
metric might be at the smallest scales.
The procedure usually taken to go from a Lorentz signature
to a Euclidean signature, the Wick rotation to imaginary time,
even if well-established in flat spacetime, 
is however not free from ambiguities in general curved spacetime
\cite{VisJ}.
Based on a consistent 
prescription for this \cite{VisJ}
(coming from reconsidering the Wick rotation as 
an analytic continuation of the metric),
results concerning the curvature tensors, 
when migrating from
Lorentzian to Euclidean signature spacetimes,
have been presented \cite{KotL},
showing it is worth pursuing in this vein 
in a qmetric context.

On the other hand,
it is not so clear
which is the role of Euclidean manifolds
at a fundamental level,
meaning 
if at the smallest scale
the reference manifold
is really to be considered Euclidean
instead of Lorentzian. 
There is, after all, no physical proof supporting
a non-Lorentzian signature for spacetime.
And the context in which
a dynamical signature change is foreseen
--from a fundamental Euclidean signature to the Lorentzian
signature we see in the universe--,
along the lines e.g.  of \cite{GreA, GreB},
has been pointed out to entail difficulties
whose origin can be traced back to how, or how strongly,
quantum field theory reacts
to changes of signature in the underlying
manifold 
(effects like
production of infinite number of particles
with infinite energy) \cite{VisAA}.
Sound arguments have been also reported,
based on the consideration 
of saddle-point approximation methods,
showing 
that quantum gravity amplitudes should be
defined first
in terms of Lorentzian path integrals \cite{SorB}.
Likewise,
from Wick rotation as analytic continuation of the metric,
the suggestion has been given that the functional integral
should be computed not over all Euclidean manifolds
but only over those compatible with a Lorentz structure \cite{VisJ}.
In causal dynamical triangulation,
evidence has been reported indeed that
one gains control on the functional integration
if the sum is not taken over
all Euclidean geometries, 
but is restricted instead
only to those 
which are associated with Lorentzian causal geometries \cite{AmbC}.

In view of all this,
one would then know if the qmetric approach
could allow to pick up quantum degrees of freedom
even never abandoning the Lorentz sector.
The aim of present study is precisely to develop
a concept of $\rho$ in the Lorentz sector directly,
i.e. with no reliance on Euclideanised space.
A partial result in this direction
has been already presented in \cite{PesK}.
There,
a notion of $\rho$ for timelike geodesics has been introduced
and its expression has been derived
(and the case of spacelike geodesics goes along similar lines).
What is left is
the consideration of null geodesics
and this is the case we try to study here.
As we will see,
this involves the introduction of a notion of quantum metric $q_{ab}$
for null separated events, this way completing
a quantum formulation of spacetime intervals.

\section{$\rho$ for timelike/spacelike geodesics}

Let us start by recalling
what we can do with timelike/spacelike geodesics.
We briefly rephrase what is reported in \cite{PesK}
for timelike case, using here a notation which encompasses
both the timelike and the spacelike case at one stroke. 
We consider timelike/spacelike geodesics
through a generic point $P$ in spacetime,
and introduce the two hypersurfaces
$\Sigma_\epsilon(P, l)$,
$\epsilon = +1$ for spacelike geodesics
and $\epsilon = -1$ for timelike ones,
of all points $p$
at assigned squared distance from $P$:

\begin{eqnarray}
  \Sigma_\epsilon(P, l) = \big\{p: \ \epsilon\sigma^2(p, P) = l^2 \big\},
  \nonumber
\end{eqnarray}
where 
$
\sigma^2(p, P)
$
is the squared geodesic distance between $P$ and $p$
($\sigma^2(p, P) = 2 \Omega(p, P)$,
with $\Omega(p, P)$ the Synge
world function \cite{Syn}),
and $l = \sqrt{l^2}$ non-negative.

Proceeding analogously
to the Euclidean definition,
$\rho$
is given in terms of generic/flat ratio of element of areas
on $\Sigma_\epsilon(P, l)$,
as measured according to the effective metric,
in the limit $l \rightarrow 0$.
For each assigned normalised vector $n^a$ at $P$
($n^a n_a = \epsilon$),
we consider the intersection point $p$ between
the geodesic $\mu(n^a)$
with tangent at $P$ $t^a(P) = n^a$
and the hypersurface $\Sigma_\epsilon(P, l)$.
Calling $y^i$, $i = 1, ..., D-1$
coordinates on $\Sigma_\epsilon(P, l)$ 
such that $y^i(p) =0$,
we consider a segment $I$ of hypersurface $\Sigma_\epsilon(P, l)$
around $p$, defined as
$
I = \{dy^i\},
$
where $dy^i$ are thought as fixed when $l$ is varied.
The $(D-1)$-dimensional area of $I$ is

\begin{eqnarray}
  d^{D-1}V(p) = \sqrt{- \epsilon h(p)} \ d^{D-1}y, \nonumber
\end{eqnarray}
where $h_{ij}$ are the components of the metric
on $\Sigma_\epsilon(P, l)$
in the coordinates $y^i$,
metric which coincides with that induced
by spacetime metric $g_{ab}$.
What we have to consider is the area $[d^{D-1}V]_q$ of $I$ 
as measured through the effective metric $q_{ab}$.

The effective metric
is described \cite{KotE, StaA}  in terms of the bitensor
$q_{ab}(p, P)$
which stems from requiring the squared
geodesic distance
$\sigma^2$
gets modified into
$
\sigma^2 \rightarrow [\sigma^2]_q =
      {S_L}(\sigma^2) 
$
with
{\bf (R1)} $S_0 = \sigma^2$,
{\bf (R2)} ${S_L}(0^\pm) = \pm L^2$,
and
{\bf (R3)}
the kernel $G(\sigma^2)$ of the d'Alembertian
gets modified into
$G(\sigma^2) \rightarrow [G]_q(\sigma^2) = G({S_L})$
in all maximally symmetric spacetimes.
These requirements give, for spacelike or timelike geodesics,
the expression

\begin{eqnarray}\label{qab}
  q_{ab}(p, P) = A(\sigma^2) g_{ab}(p) +
  \epsilon \Big(\frac{1}{\alpha(\sigma^2)} - A(\sigma^2)\Big) t_a(p) t_b(p),
\end{eqnarray}
where $t^a$ is the normalized tangent vector
($g_{ab} t^a t^b = \epsilon$),
not going to change its timelike or spacelike character
when in the qmetric,

\begin{eqnarray}\label{A}
  A = \frac{S_L}{\sigma^2}
  \Big(\frac{\Delta}{\Delta_S}\Big)^\frac{2}{D-1},
\end{eqnarray}

\begin{eqnarray}\label{alpha}
  \alpha = \frac{S_L}{\sigma^2 (S^\prime_L)^2}
\end{eqnarray}
($^\prime$ indicates differentiation with respect
to the argument $\sigma^2$),
where
\begin{eqnarray}\label{vanVleck}
  \Delta(p, P) = - \frac{1}{\sqrt{g(p) g(P)}}
  {\rm det}\Big[-\nabla^{(p)}_a \nabla^{(P)}_b \frac{1}{2} \sigma^2(p, P)\Big]
\end{eqnarray}
is the van Vleck determinant
(\cite{vVl, Mor, DeWA, DeWB}; see \cite{Xen, VisA, PPV})
which is a biscalar,
and the biscalar $\Delta_S(p, P)$ is
$\Delta_S(p, P) = \Delta({\tilde p}, P)$,
where $\tilde p$ is
that point on the geodesic through $P$ and $p$
(on the same side of $p$ with respect to $P$)
which has $\sigma^2({\tilde p}, P) = S_L(p, P)$.
$\alpha$ is determined by the request
that the formula for squared geodesic distance

\begin{eqnarray}\label{HJ}
  g^{ab} \partial_a\sigma^2 \partial_b\sigma^2 = 4 \sigma^2
\end{eqnarray}
(Hamilton-Jacobi equation)
gets transformed into
$q^{ab} \partial_a S_L \partial_b S_L = 4 S_L$;
$A$ by the request {\bf R3}.

From
the effective metric $[h_{ab}]_q(p, P)$ induced by $q_{ab}(p, P)$
at $p$ on $\Sigma_\epsilon(P, l)$,
we get 
the effective-metric $(D-1)$-dimensional area of $I$ as
\begin{eqnarray}
 [d^{D-1}V]_q(p, P)
 = \Big[\sqrt{- \epsilon h}\Big]_q(p, P) \ d^{D-1}y . \nonumber
\end{eqnarray}
As in the Euclidean approach,
$\rho$ can then be defined as the ratio
of effective-metric $(D-1)$-dimensional area of $I$
for the actual metric configuration, $[d^{D-1}V]_{q(g)}(p, P)$, 
to what we would have were spacetime flat, $[d^{D-1}V]_{q(\eta)}(p, P)$
($\eta_{ab}$ is Minkowski metric), in the
limit $p \rightarrow P$ along $\mu(n^a)$,
i.e.

\begin{eqnarray}\label{rho_def}
  \rho(P, n^a) =
  \bigg(\lim_{p \rightarrow P}
  \frac{[d^{D-1}V]_{q(g)}(p, P)}{[d^{D-1}V]_{q(\eta)}(p, P)}\bigg)
  _{\mu(n^a)} .
\end{eqnarray}

$\rho$ is then derived
in terms of the quantities $A$ and $\alpha$ defining the effective metric.
The effective metric $[h_{ab}]_q$ induced by $q_{ab}$
turns out to be

\begin{eqnarray}
  [h_{ab}]_q(p, P) = A(\sigma^2) h_{ab}(p)
  \nonumber
\end{eqnarray}  
\cite{KotG},
which implies

\begin{eqnarray}
  \Big[\sqrt{- \epsilon h}\Big]_q(p, P) = A(\sigma^2)^{\frac{D-1}{2}}
  \sqrt{- \epsilon h(p)},
  \nonumber
\end{eqnarray}  
and then

\begin{eqnarray}
  [d^{D-1}V]_q(p, P)
  &=& A(\sigma^2)^{\frac{D-1}{2}} d^{D-1}V(p), \nonumber
\end{eqnarray}
where $d^{D-1}V(p)$ indicates the proper area of $I$
according to the ordinary metric.
Here we see that
only $A$, and not $\alpha$,
is actually involved
in the determination of $\rho$.

Introducing
on $\Sigma_\epsilon$, in a neighbourhood of $p$,
mutually orthogonal coordinates $z^i$ 
such that,
chosen any one of them, $z^{\bar i}$,
it can be written in the form
$z^{\bar i} = l \eta$
with the parameter $\eta$ such that $l d\eta$ is proper distance
or proper-time difference,
and chosing as $I$ the (hyper)cube ${dz^i}$
defined by $dz^i = l d\eta, \forall i$,
we obtain

\begin{eqnarray}
  [d^{D-1}V]_q(p, P)
  &=& A(\sigma^2)^{\frac{D-1}{2}} l^{D-1}
      \big(1 + {\cal O}(l^2)\big) (d\eta)^{D-1}
      \nonumber
\end{eqnarray}
where the ${\cal O}(l^2)$ term
represents the effects of curvature
(and is thus of course absent in flat case),
and clearly
$l = \sqrt{\epsilon \sigma^2}.$
Using the expression (\ref{A}) for $A$,
we get

\begin{eqnarray}
  [d^{D-1}V]_q(p, P)
  = [\epsilon S_L]^{\frac{D-1}{2}}
      \frac{\Delta(p, P)}{\Delta_S(p, P)} \
      \big(1 + {\cal O}(l^2)\big) (d\eta)^{D-1}   \nonumber
\end{eqnarray}
and,
in the limit $p \rightarrow P$ along $\mu(n^a)$,

\begin{eqnarray}
  \lim_{l \rightarrow 0}
  \ [d^{D-1}V]_q(p, P)
  = L^{D-1} \frac{1}{\Delta_L(P, n^a)} (d\eta)^{D-1},     \nonumber
\end{eqnarray}  
with
$\Delta_L(P, n^a) = \Delta({\bar p}, P)$, where $\bar p$ 
is that point on geodesic $\mu(n^a)$ (on the side in the direction $n^a$)
which has $l = L$.

This shows that both the numerator and the denominator
in expression (\ref{rho_def}) remain non vanishing
in the coincidence limit $p \rightarrow P$,
exactly as it happens in Euclidean case.
Since for flat spacetime $\Delta = 1$ identically and then also
$\Delta_L = 1$,
we have finally

\begin{eqnarray}\label{rho}
  \rho(P, n^a)
  = \frac{1}{\Delta_L(P, n^a)},
\end{eqnarray}
where the $\Delta_L$ is that of generic metric $g_{ab}$. 
The scope of this exact expression for $\rho$
clearly includes strictly Riemannian manifolds (as that from Euclideanisation). 

Expanding $\Delta(p, P)$ in powers of $l$
(\cite{DeWA}; \cite{Xen, VisA, PPV}),

\begin{eqnarray}\label{expansion}
  \Delta(p, P) = 1 + \frac{1}{6} l^2 R_{ab} t^a t^b
  + o\big(l^2 R_{ab} t^a t^b\big),
\end{eqnarray}  
($t^a t_a =\epsilon)$ 
gives

\begin{eqnarray}\label{DeltaL}
 \Delta_L(P, n^a) =  
 1 + \frac{1}{6} L^2 R_{ab}(P) n^a n^b + o\big(L^2 R_{ab}(P) n^a n^b\big),
\end{eqnarray}
and 

\begin{eqnarray}\label{rho_expansion}
  \rho(P, n^a) =
  1 -\frac{1}{6} L^2 R_{ab}(P) n^a n^b + o\big(L^2 R_{ab}(P) n^a n^b\big).
\end{eqnarray}
Again, this identically applies also to Riemannian manifolds
(as that from Euclideanisation),
and its form coincides with the expansion obtained
\cite{Pad04, Pad06, Pad08, Pad10} defining $\rho$
in the Euclideanised space.

\section{qmetric and null geodesics}

If we try to extend the scope of effective metric approach
to include null geodesics,
we have that expression (\ref{qab}) becomes ill defined in this case
since
$\sigma^2 =0$ all along any null geodesic,
and in principle we are then in trouble.
We notice however the following.
Any affine parametrization $\lambda$ of a null geodesic can be thought of
as a measure of distance along the geodesic performed
by a canonical observer
picked up at a certain point $x$ of the geodesic and parallel transported
along the geodesic.
Since, when going to the effective metric $q_{ab}$,
the squared distance in the coincidence limit is the finite value
$\epsilon L^2$ (request {\bf R2} above),
we could expect the effect of the qmetric in the null case is
to induce a mapping of the parametrization $\lambda$ to a new parametrization
$\tilde \lambda = \tilde \lambda(\lambda)$,
with ${\tilde \lambda} \rightarrow L$ when
$\lambda(p, P) \rightarrow 0$, i.e. when $p \rightarrow P$.
In analogy with the spacelike/timelike case,
we can then think to give
an expression for $q_{ab}(p, P)$ when $p$ is on a null
geodesic from $P$
in terms of two functions $\alpha_\gamma = \alpha_\gamma(\lambda)$
and $A_\gamma = A_\gamma(\lambda)$ defined on the geodesic,
and determined by a condition on the squared geodetic distance
and on the d'Alembertian.
In other words,
this suggests we assume
that the effects of the existence of a limiting length  
are captured by an effective metric bitensor $q_{ab}$ as above,
with its expression on a null geodesic stemming from requiring
the affine parametrization $\lambda$ gets modified
into $\lambda \rightarrow [\lambda]_q = {\tilde \lambda}(\lambda)$
with
{\bf (G1)}
${\tilde \lambda} = \lambda$ if $L = 0$
(or ${\tilde \lambda} \simeq \lambda$ when $\lambda \rightarrow \infty$),
{\bf (G2)}
${\tilde \lambda}(0^+) = L$,
and
{\bf (G3)}
the kernel $G(\sigma^2)$ gets modified into $[G]_q(\sigma^2) = G(S_L)$
in all maximally symmetric spacetimes, i.e {\bf (G3)} coincides
with {\bf (R3)} above on null geodesics.

We see that dealing with the null case appears quite not
so obvious,
in that we are forced to rewrite for this case from scratch
the rules to go to the qmetric given a metric, in terms of an
affine parameter $\lambda$ defined on null geodesics only,
i.e. $q_{ab}$ is defined strictly on null geodesics
and knows nothing outside them.
And this, morover,
leads to the tricky circumstances
that the operators we look at when constraining the expression
for $q_{ab}$ (e.g. the d'Alembertian) should be considered in a form
which does not hinge on any knowledge,
regarding the elements which enter the definition of the operator itself
(directional derivatives, vectors),
of what happens outside the $(D-1)$-dimensional submanifold
swept by all the null geodesics emanating from a point.

Let $\gamma$ be a null geodesic through $P$,
with affine parameter
$\lambda = \lambda(p, P)$ with $\lambda(P, P) = 0$,
and null tangent vector $l^a = \frac{dx^a}{d\lambda}$,
i.e. $\nabla_a (\sigma^2) = 2 \lambda l_a$ (see e.g. \cite{PPV}). 
We introduce a canonical observer at $P$, with velocity $V^a$,
such that $l_a V^a = -1$. By parallel transport of the observer
along $\gamma$, this relation extends all along $\gamma$,
with $\lambda$ having the meaning of a distance
as measured by this observer.
We affinely parametrize any other null geodesic $\hat \gamma$
which goes through $P$, and require ${\hat l}_a V^a = -1$.
What we obtain this way,
is
a $(D-1)$-dimensional congruence $\Gamma$ of null geodesics
emanating from $P$ which is affinely parametrized
and has deviation vectors orthogonal to the geodesics.
We introduce a second null vector $m^a$ at $P$, defined by
$m^a \equiv 2 V^a - l^a$,
parallelly transported along the geodesic.
This gives $m_a V^a = -1$ and $m_a l^a = -2$ all along $\gamma$.
The vector $m^a$ does depend on the observer we have chosen.

Let $q_{ab}(p, P)$, $p$ on $\gamma$, be of the form

\begin{eqnarray}\label{qab_null}
  q_{ab} = A_\gamma g_{ab} -\frac{1}{2}
          \Big(\frac{1}{\alpha_\gamma} - A_\gamma\Big) (l_a m_b + m_a l_b).
\end{eqnarray}
From $q^{ab} q_{bc} = \delta^a_c$, we get

\begin{eqnarray}
  q^{ab} = \frac{1}{A_\gamma} g^{ab} + \frac{1}{2}
          \Big(\frac{1}{A_\gamma} - \alpha_\gamma\Big) (l^a m^b + m^a l^b),
\end{eqnarray}
where $l^a = g^{ab} l_b$, $m^a = g^{ab} m_b$. 
Notice that $q^{ab} l_a l_b = 0$, and the geodesic is null also according
to the qmetric.

Our first task is to determine the form of $\alpha_\gamma$.
To this aim,
we use of the request that
$[l^a]_q = dx^a/d\tilde\lambda$
be parallelly transported according to the qmetric.
We need this,
if $\tilde\lambda$ has to be interpreted
as a (quantum) arc-length according to a canonical observer.
We have

\begin{eqnarray}
  [l^b]_q \ [\nabla_b]_q \ [l_c]_q
  &=&
  \frac{d\lambda}{d\tilde\lambda} \, l^b
  \Bigg(\partial_b\bigg(\frac{d\lambda}{d\tilde\lambda}
  \frac{1}{\alpha_\gamma} l_c\bigg)
  - [\Gamma^a_{bc}]_q
  \frac{d\lambda}{d\tilde\lambda} \ \frac{1}{\alpha_\gamma} l_a\Bigg),  
\end{eqnarray}
where
$
[l_c]_q
=
q_{ac} [l^a]_q
=
\frac{d\lambda}{d\tilde\lambda} \,
\frac{1}{\alpha_\gamma} \, l_c.
$
Here,
from
$
{\Gamma^a}_{bc} = \frac{1}{2} g^{ad}
(-\partial_d g_{bc} + \partial_c g_{bd} +\partial_b g_{dc}),
$
we have

\begin{eqnarray}
  [{\Gamma^a}_{bc}]_q
  &=&
  \frac{1}{2} q^{ad}
  (-\partial_d q_{bc} + \partial_c q_{bd} +\partial_b q_{dc})
  \nonumber \\
  &=& \frac{1}{2} q^{ad}
  (-\nabla_d q_{bc} + 2 \nabla_{\left(b\right.} q_{\left.c\right)d}) + {\Gamma^a}_{bc}
  \nonumber
\end{eqnarray}
(cf. \cite{KotG}).
Using of this, we get

\begin{eqnarray}\label{rayqab2_23_1}
  [l^b]_q \ [\nabla_b]_q \ [l_c]_q
  &=&
  \frac{d\lambda}{d\tilde\lambda} \, l_c \,
  \frac{d}{d\lambda} \bigg(\frac{d\lambda}{d\tilde\lambda} \,
  \frac{1}{\alpha_\gamma}\bigg)
  - \frac{1}{2} \, \bigg(\frac{d\lambda}{d\tilde\lambda}\bigg)^2 \,
  l^d l^b \, (-\nabla_d q_{bc} + 2 \nabla_{\left(b\right.} q_{\left.c\right)d})
  \nonumber \\
  &=&
  \frac{d\lambda}{d\tilde\lambda} \, l_c \,
  \frac{d}{d\lambda} \bigg(\frac{d\lambda}{d\tilde\lambda} \,
  \frac{1}{\alpha_\gamma}\bigg)
  - \bigg(\frac{d\lambda}{d\tilde\lambda}\bigg)^2 \,
  \Big(\frac{1}{\alpha_\gamma} - A_\gamma\Big) \
  l^b \nabla_c l_b,
\end{eqnarray}  
where
in the 1st equality we used of
$
l^b \nabla_b l_c = 0
$
and of
$
q^{ad} l_a =
\alpha_\gamma l^d,
$
and, in the 2nd, of
$
l^d l^b \nabla_c q_{bd}
=
2 \big(\frac{1}{\alpha_\gamma} - A_\gamma\big) \, l^b \nabla_c l_b.
$
Here,
$
\nabla_c l_b
$
brings to consider variations of $l_b$ outside $\Gamma$.
However,
in whichever way might $l^b$ null be thougth to be extended outside $\Gamma$,
always it will hold true that
$
\nabla_c(l^b l_b) = 2 l^b \nabla_c l_b = 0.
$
We have then

\begin{eqnarray}\label{rayqab2_23_2}
  [l^b]_q \ [\nabla_b]_q \ [l_c]_q
  &=&
  \frac{d\lambda}{d\tilde\lambda} \, l_c \,
  \frac{d}{d\lambda} \bigg(\frac{d\lambda}{d\tilde\lambda} \,
  \frac{1}{\alpha_\gamma}\bigg). 
\end{eqnarray}  
$
[l^b]_q \ [\nabla_b]_q \ [l_c]_q = 0
$
requires
$
\alpha_\gamma = K \frac{d\lambda}{d\tilde\lambda},
$
with $K$ a constant.
To determine $K$ we use the following.
When
$\lambda\to \infty$,
$d\lambda/d\tilde\lambda \to 1$
and
we must have also
$
q_{ab} \to g_{ab}.
$
This implies both
$
\lim_{\lambda\to \infty} A_\gamma = 1
$
and
$
K = 1.
$
What we get is thus

\begin{eqnarray}\label{alpha_gamma}
  \alpha_\gamma = \frac{1}{d{\tilde \lambda}/d\lambda}. 
\end{eqnarray}

As for the determination of $A_\gamma$,
we have to refer to {\bf G3},
i.e we consider the d'Alembertian
in maximally symmetric spaces
at points on null geodesics.
What we try first, is to find out some convenient expression for the
d'Alembertian.
Due to maximal symmetry,
we can think in terms of $f = f(\sigma^2)$
and write

\begin{eqnarray}
  \Box f &=& \nabla_a \nabla^a f
  \nonumber \\
  &=& \nabla_a\Big(\partial^a\sigma^2 \frac{df}{d\sigma^2}\Big)
  \nonumber \\
  &=& \big(\nabla_a  \partial^a\sigma^2\big) \frac{df}{d\sigma^2} +
  \big(\partial^a\sigma^2\big) \partial_a \frac{df}{d\sigma^2}
  \nonumber \\
  &=& \big(\nabla_a  \partial^a\sigma^2\big) \frac{df}{d\sigma^2} +
  \big(\partial^a\sigma^2\big)
  \big(\partial_a\sigma^2\big) \frac{d^2f}{d(\sigma^2)^2}.
  \nonumber
\end{eqnarray}
When going to null geodesic $\gamma$,
$ \big(\partial^a\sigma^2\big) \big(\partial_a\sigma^2\big)
\rightarrow (2 \lambda l^a) (2 \lambda l_a) = 0 $
and we get

\begin{eqnarray}
  \nonumber
  \Box f = \big(\nabla_a  \partial^a\sigma^2\big) \frac{df}{d\sigma^2}.
\end{eqnarray}
At a point $p'$ close to $\Gamma$
but, possibly, not exactly on it,
we can write (cf. \cite{VisA})

\begin{eqnarray}\label{notquiteongamma}
  \partial^a\sigma^2_{|p'} =
    2 \lambda \ l^a_{|p'} + 2 \nu \ m^a_{|p'}, 
\end{eqnarray}
where $\lambda$ and $\nu$ are curvilinear null coordinates of $p'$
(there is a unique point $p$ on $\Gamma$ from which $p'$
is reachable
through a null geodesic $\beta$ with tangent $m^a$ at $p$;
$\nu$ is the affine parameter of $p'$ along $\beta$,
with $\nu(p) = 0$),
$l^a_{|p'}$ and $m^a_{|p'}$ are $l^a$ and $m^a$ parallel
transported along $\beta$ from $p$ to $p'$.
This gives, on $\gamma$,

\begin{eqnarray}\label{nabla_partial}
  \nabla_a\partial^a\sigma^2
  &=& 2 \ \big(\lambda \nabla_a l^a
  + l^a \partial_a\lambda + m^a \partial_a\nu\big)
  \nonumber \\
  &=& 2 \ \big(\lambda \nabla_a l^a + 2),
\end{eqnarray}
and then

\begin{eqnarray}
  \Box f
  =
  \big(4 + 2 \lambda \nabla_a l^a\big) \frac{df}{d\sigma^2}
  =
  \big(4 + 2 \lambda \nabla_i l^i\big) \frac{df}{d\sigma^2}, 
\end{eqnarray}
$i = 1, ..., D-1$ indices of components on $\Gamma$.
Here, we emphasized the fact that,
since the covariant derivative of $l^a$ along $\beta$ is 0,
$\nabla_a l^a$
is completely defined within $\Gamma$
and coincides with the expansion of $\Gamma$,
$\nabla_a l^a = \nabla_i l^i$.

Going to the qmetric,
the geodesic $\gamma$ remains null,
and we have

\begin{eqnarray}\label{Box_q1}
  [\Box f]_q &=& \big(4 + 2 [\lambda \nabla_a l^a]_q\big)
  \Big[\frac{df}{d\sigma^2}\Big]_q
  \nonumber \\
  &=& \big(4 + 2 [\lambda]_q \ [\nabla_a l^a]_q\big)
  \frac{d[f]_q}{dS_L}
  \nonumber \\
  &=& \big(4 + 2 {\tilde \lambda}
  \ [\nabla_a l^a]_q\big)
  \Big(\frac{df}{d\sigma^2}\Big)_{|\sigma^2=S_L}.
\end{eqnarray}
Here
$
[l^a]_q = dx^a/d{\tilde \lambda} = (d\lambda/d{\tilde \lambda}) \ l^a,
$
and
$
f\!\!: \sigma^2 \mapsto f(\sigma^2)
$
gets mapped by the qmetric into
$
[f]_q\!\!: \sigma^2 \mapsto S_L \mapsto f(S_L)=[f]_q(\sigma^2)
$
which has
$
\frac{d[f]_q}{dS_L} = (\frac{df}{d\sigma^2})_{|\sigma^2=S_L}.
$
As for the divergence,
we have
$
[\nabla_a l^a]_q =
[(\partial_a + {\Gamma^b}_{ab}) l^a]_q.
$
From

\begin{eqnarray}
  [{\Gamma^b}_{ab}]_q
  &=&
  \frac{1}{2} q^{bc}
  (-\nabla_c q_{ab} + 2 \nabla_{\left(a\right.} q_{\left.b\right)c}) + {\Gamma^b}_{ab}
  \nonumber \\
  &=&
  \frac{1}{2} q^{bc} \nabla_a q_{bc} + {\Gamma^b}_{ab},
  \nonumber
\end{eqnarray}
we get

\begin{eqnarray}
  \nonumber
  [\nabla_a l^a]_q =
  \nabla_a\Big(\frac{d\lambda}{d{\tilde \lambda}} l^a\Big)
  + \frac{1}{2} q^{bc} (\nabla_a q_{bc})
  \ \frac{d\lambda}{d{\tilde\lambda}} \ l^a.
\end{eqnarray}
This expression openly shows that
all differentials are indeed taken on $\Gamma$. 
Using formula (\ref{qab_null}) for $q_{ab}$,
direct computation gives

\begin{eqnarray}\label{rayqab2_25_1}
  [\nabla_a l^a]_q
  &=& \frac{d\lambda}{d{\tilde\lambda}} \ \nabla_i l^i +
  \frac{d}{d\lambda} \Big(\frac{d\lambda}{d{\tilde\lambda}}\Big) +
  \frac{1}{2} \frac{d\lambda}{d{\tilde\lambda}}
  \Big\{ (D-2) \frac{d}{d\lambda} \ln A_\gamma -
  2 \frac{d}{d\lambda} \ln \alpha_\gamma\Big\}
  \nonumber \\
  &=&\frac{d\lambda}{d{\tilde\lambda}} \ \nabla_i l^i
  +
  \frac{1}{2} (D-2) \frac{d\lambda}{d{\tilde\lambda}}
  \frac{d}{d\lambda} \ln A_\gamma,
  \nonumber
\end{eqnarray}  
where, in the 2nd equality, use of the expression (\ref{alpha_gamma})
for $\alpha_\gamma$ was made.
Inserting this into equation (\ref{Box_q1}),
we get

\begin{eqnarray}
  [\Box f]_q =
  \Big\{4 + 2 {\tilde\lambda} \frac{d\lambda}{d{\tilde\lambda}} \ \nabla_i l^i
  +
  {\tilde\lambda} \ (D-2) \frac{d\lambda}{d{\tilde\lambda}}
  \frac{d}{d\lambda} \ln A_\gamma\Big\}
  \Big(\frac{df}{d\sigma^2}\Big)_{|\sigma^2=S_L}.       
\end{eqnarray}

Now we are ready to implement condition {\bf G3}.
We require that,
if $G = G(\sigma^2_{|\tilde p'})$,
with $\sigma^2_{|\tilde p'} = S_L$, 
is solution to $\Box G = 0$
in $\tilde p$ at $\tilde\lambda$ on $\gamma$
($\tilde p'$ is in a ($D$-dim) neighbourhood
of $\tilde p$;
with
$
\sigma^2_{|\tilde p'}/\sigma^2_{|p'} \to \tilde\lambda^2/\lambda^2
$
when
$
\tilde p' \to \tilde p
$
and
$
p' \to p,
$
due to continuity reasons),
i.e. if
$\Box G_{|\tilde p} = 0$,
then
$[G]_q(\sigma^2) \equiv G(S_L(\sigma^2))$
be solution of
$
[\Box G]_q = 0
$
in $p$ at $\lambda$ on $\gamma$,
i.e.

\begin{eqnarray}\label{tra21e22}
  4 + 2 {\tilde\lambda} \frac{d\lambda}{d{\tilde\lambda}} \ \nabla_i l^i
  +
  {\tilde\lambda} \ (D-2) \frac{d\lambda}{d{\tilde\lambda}}
  \frac{d}{d\lambda} \ln A_\gamma = 0        
\end{eqnarray}
in $p$.

We proceed first
to calculate
$\Box G_{|\tilde p}$.
In $\tilde p'$, we have

\begin{eqnarray}
  \Box G_{|\tilde p'}
  &=&
  \big(\nabla_a \nabla^a G\big)_{|\tilde p'}
  \nonumber \\
  &=&
  \nabla_a \Big( (\partial^a \sigma^2_{|\tilde p'})
  \frac{dG}{d\sigma^2_{|\tilde p'}}\Big)
  \nonumber \\
  &=&
  \big(\nabla_a\partial^a \sigma^2_{|\tilde p'}\big)
  \frac{dG}{d\sigma^2_{|\tilde p'}}
  + \big(\partial^a \sigma^2_{|\tilde p'}\big) \,
  \partial_a \frac{dG}{d\sigma^2_{|\tilde p'}}
  \nonumber \\
  &=&
  \big(\nabla_a\partial^a \sigma^2_{|\tilde p'}\big)
  \frac{dG}{d\sigma^2_{|\tilde p'}} +
  \big(\partial^a \sigma^2_{|\tilde p'}\big)
  \big(\partial_a \sigma^2_{|\tilde p'}\big) \,
  \frac{d}{d\sigma^2_{|\tilde p'}}
  \bigg(\frac{dG}{d\sigma^2_{|\tilde p'}}\bigg).
\end{eqnarray}
When
$\tilde p' \to \tilde p$ on $\gamma$,
$
\big(\partial^a \sigma^2_{|\tilde p'}\big)
\big(\partial_a \sigma^2_{|\tilde p'}\big)
\to
(2 {\tilde\lambda} \, {l^a}_{|{\tilde p}})
(2 {\tilde\lambda} \, {l_a}_{|{\tilde p}})
= 0
$
and thus what matters here is the first term. 
We have

\begin{eqnarray}
  \nabla_a \partial^a \sigma^2_{|\tilde p'}
  &=&
  \big(\nabla_a (2 \lambda \, l^a + 2 \nu \, m^a)\big)_{|\tilde p'}
  \nonumber \\
  &=&
  \nabla_a(2 \tilde\lambda \, {l^a}_{|\tilde p'}
  + 2 \tilde\nu \, {m^a}_{|\tilde p'}),
\end{eqnarray}  
where we used of 
relation (\ref{notquiteongamma})
and
wrote
$
\tilde\lambda = \frac{1}{2} ({\tilde t} + {\tilde r}),
$
$
\tilde\nu = \frac{1}{2} ({\tilde t} - {\tilde r}).
$
When going to $\gamma$,
we get

\begin{eqnarray}
  (\nabla_a \partial^a \sigma^2)_{|\tilde p}
  &=&
  2 \tilde\lambda \, (\nabla_a l^a)_{|\tilde p}
  + 2 {l^a}_{|\tilde p} \, \nabla_a \tilde\lambda
  + 2 {m^a}_{|\tilde p} \, \nabla_a \tilde\nu
  \nonumber \\
  &=&
  2 \tilde\lambda \, (\nabla_i l^i)_{|\tilde p}
  + 4,
\end{eqnarray}
for $\tilde \lambda$ is 
the affine parameter $\lambda$ at $\tilde p$,
and
$
{m^a}_{|\tilde p}
=
dx^a/d{\tilde\nu}.
$
Thus,
we have

\begin{eqnarray}\label{qmetric2_27_3}
  \Box G_{|\tilde p}
  &=&
  \bigg(
  2 \tilde\lambda \, (\nabla_i l^i)_{|\tilde p}
  + 4 \bigg) \,
  \frac{dG}{d\sigma^2_{|\tilde p}}.
\end{eqnarray}
$
\Box G_{|\tilde p} = 0
$
then means

\begin{eqnarray}
  2 \tilde\lambda \, (\nabla_i l^i)_{|\tilde p}
  + 4
  = 0.
\end{eqnarray}
Inserting this into (\ref{tra21e22}),
one obtains

\begin{eqnarray}
  -2 \tilde\lambda \,
  (\nabla_i l^i)_{|\tilde p}
  +
  2 \tilde\lambda \, \frac{d\lambda}{d\tilde\lambda} \,
  \nabla_i l^i
  +
  \tilde\lambda \, (D-2) \, \frac{d\lambda}{d\tilde\lambda} \,
  \frac{d}{d\lambda} \ln A_\gamma
  = 0,
  \nonumber
\end{eqnarray}
which is

\begin{eqnarray}\label{44_5}
  -2 \, \frac{d\tilde\lambda}{d\lambda} \, (\nabla_i l^i)_{|\tilde p}
  + 
  2  \, \nabla_i l^i
  +
  (D-2) \, \frac{d}{d\lambda} \ln A_\gamma
  = 0.
\end{eqnarray}

Thanks to the relation
(\cite{DeWA, DeWB}; see \cite{VisA, PPV})

\begin{eqnarray}
  \nabla_a^{(p)}\big[\Delta(p, P) \nabla^a_{(p)} \sigma^2(p, P)\big] =
  2 D \ \Delta(p, P)
\end{eqnarray}
(valid for spacelike/timelike as well as null geodesics),
which gives

\begin{eqnarray}
  \nonumber
  \nabla_a \partial^a \sigma^2 =
  2 D + (\nabla_a \ln \Delta^{-1}) \ \partial^a \sigma^2
\end{eqnarray}
with $\partial^a \sigma^2 = 2 \lambda l^a$ on $\gamma$,
using (\ref{nabla_partial})
the expansion of the congruence 
can be usefully expressed in terms of the
van Vleck determinant as (cf. \cite{VisA})

\begin{eqnarray}
  \nabla_a l^a =
  \nabla_i l^i =
  \frac{D-2}{\lambda} + \frac{d}{d\lambda}\ln \Delta^{-1}
\end{eqnarray}
and

\begin{eqnarray}
  (\nabla_a l^a)_{|\tilde p} =
  (\nabla_i l^i)_{|\tilde p} =
  \frac{D-2}{\tilde\lambda} +
  \frac{d}{d\tilde\lambda}\ln \Delta_S^{-1},
\end{eqnarray}
where $\Delta_S$ is the van Vleck determinant evaluated at $\tilde p$.

Substituting this, equation (\ref{44_5}) above
becomes

\begin{eqnarray}
  -2 \, \bigg( \frac{d\tilde\lambda}{d\lambda} \, \frac{D-2}{\tilde\lambda} +
  \frac{d}{d\lambda}\ln \Delta_S^{-1}\bigg)
  +
  2 \, \bigg(\frac{D-2}{\lambda} + \frac{d}{d\lambda}\ln \Delta^{-1}\bigg)
  +
  (D-2) \, \frac{d}{d\lambda} \ln A_\gamma
  = 0,
  \nonumber
\end{eqnarray}
or

\begin{eqnarray}
  -2 \, \frac{d\tilde\lambda}{d\lambda} \, \frac{1}{\tilde\lambda}
  -\frac{2}{D-2} \, \frac{d}{d\lambda}\ln \Delta_S^{-1}
  +\frac{2}{\lambda}
  +\frac{2}{D-2} \, \frac{d}{d\lambda}\ln \Delta^{-1}
  + \frac{d}{d\lambda} \ln A_\gamma
  = 0,
  \nonumber
\end{eqnarray}
which is

\begin{eqnarray}
  \frac{d}{d\lambda} \ln
  \Bigg(\frac{\lambda^2}{{\tilde\lambda}^2} \,
  \Big(\frac{\Delta_S}{\Delta}\Big)^{\frac{2}{D-2}}
  A_\gamma\Bigg) = 0.  
\end{eqnarray}
Thus

\begin{eqnarray}
  A_\gamma =
  C \, \frac{\tilde\lambda^2}{{\lambda}^2} \,
  \Big(\frac{\Delta}{\Delta_S}\Big)^{\frac{2}{D-2}},
  \nonumber
\end{eqnarray}  
where $C$ is a constant.
To determine $C$,
we note that
using this expression
we get, 
in the $\lambda \rightarrow \infty$ limit,
$A_\gamma \rightarrow C$.
Since,
as we saw,
$q_{ab} \rightarrow g_{ab}$ in the same limit
implies
$A_\gamma \rightarrow 1$,
we get
$C = 1$.
Our expression for $A_\gamma$ is finally

\begin{eqnarray}\label{A_gamma}
  A_\gamma =
  \frac{\tilde\lambda^2}{{\lambda}^2} \,
  \Big(\frac{\Delta}{\Delta_S}\Big)^{\frac{2}{D-2}}.
\end{eqnarray}
In conclusion,
what we have got in this Section
is
the expression (\ref{qab_null})
for the qmetric $q_{ab}$ for null geodesics,
with the functions $\alpha_\gamma$ and $A_\gamma$ in it, defined
on the null geodesics, required to have the expressions given
by equations (\ref{alpha_gamma}) and (\ref{A_gamma}).
We notice that
no dependence on the chosen canonical
observer is present in
$\alpha_\gamma$ or $A_\gamma$.
The expression (\ref{qab_null})
for $q_{ab}$, however,
does depend on the observer, through $m^a$.

\section{$\rho$ for null geodesics (Lorentz sector)}

Using the results of previous Section,
let us proceed now to try to find out an expression for $\rho$ for
null geodesics.
In complete analogy with the timelike/spacelike case,
this quantity can be defined, in the Lorentz sector, as
(cf. equation (\ref{rho_def}))

\begin{eqnarray}\label{rho_null}
    \rho(P, l^a) =
  \bigg(\lim_{p \rightarrow P}
  \frac{[d^{D-1}V]_{q(g)}(p, P)}{[d^{D-1}V]_{q(\eta)}(p, P)}\bigg)
  _{\gamma(l^a)}.
\end{eqnarray}    
Here,
$\gamma(l^a)$ is a
null geodesic through $P$,
affinely parameterized through $\lambda = \lambda(p, P)$ with
$\lambda(P, P) = 0$,
with tangent vector $k^a = dx^a/d\lambda$ along it
which takes the value
$l^a$ at $P$, i.e. $l^a = k^a_{|P}$.
The limit is taken for $p$ approaching $P$ along $\gamma(l^a)$.
$
d^{D-1}V
$
is 
a $(D-1)$-dim volume element of a null hypersurface $\Sigma_\gamma$ through $p$,
defined by $\Phi = {\rm const}$, with $-(\partial_a \Phi)_{|p} = (k_a)_{|p}$. 
Apart from this condition on the gradient,
the hypersurface $\Sigma_\gamma$ is arbitrary.
$
[d^{D-1}V]_q
$
is the volume of that same element of hypersurface,
according to the qmetric,
with
$\Sigma_\gamma$ being null also according to the qmetric
($q^{ab} k_a k_b = 0$, as we saw before).
The index $q(g)$, or simply $q$, refers to a generic metric $g_{ab}$,
while $q(\eta)$ is for the flat case.

$d^{D-1}V$ can be written as follows
(\cite{PoiA, PadN}, e.g.).
Using the vector $m^a$ as defined in the previous Section,
we can write
the metric transverse to $k^a$ at $p$
as

\begin{eqnarray}
  \nonumber
  h_{ab} = g_{ab} + \frac{1}{2} (k_a m_b + m_a k_b).
\end{eqnarray}
Introducing the coordinates $(\lambda, \theta^A)$ for $\Sigma_\gamma$,
with the coordinates $\theta^A$ spanning the $(D-2)$-dim space
transverse to the generators of $\Sigma_\gamma$,
we have the induced metric on the $(D-2)-$dim space is given by

\begin{eqnarray}
  \nonumber
  \sigma_{AB}
  &=& g_{ab} e^a_A e^b_B
  \nonumber \\
  &=& h_{ab} e^a_A e^b_B
  \nonumber
\end{eqnarray}
in terms of the vectors
$e^a_A = \big(\frac{\partial x^a}{\partial \theta^A}\big)_\lambda$
($e^a_A$ is orthogonal to both $k^a$ and $m^a$).  
The volume element can then be written as

\begin{eqnarray}
d^{D-1}V = \sqrt{\sigma} \ d^{D-2}\theta \ d\lambda, 
\end{eqnarray}
with $\sigma = \det (\sigma_{AB})$.

Going to the qmetric,
$k^a = dx^a/d\lambda$
gets mapped to
$[k^a]_q = dx^a/d{\tilde\lambda} = (d\lambda/d{\tilde\lambda}) k^a$.
$\Sigma_\gamma$ is null also according to the qmetric,
and the metric transverse (according to $q_{ab}$) to $[k^a]_q$
is given by

\begin{eqnarray}
  [h_{ab}]_q = q_{ab} +
  \frac{1}{2} \Big([k_a]_q [m_b]_q + [m_a]_q [k_b]_q\Big),
  \nonumber
\end{eqnarray}  
with
$
[k_a]_q
= q_{ab} [k^a]_q
= \frac{1}{\alpha_\gamma} \frac{d\lambda}{d{\tilde\lambda}} k_a
= k_a,
$
and
$
[m_a]_q
= \frac{d{\tilde\lambda}}{d\lambda} m_a
$
(to get
$q^{ab} [k_a]_q [m_b]_q = -2$).  
Using the expression (\ref{qab_null}) for $q_{ab}$,
we get

\begin{eqnarray}
  [h_{ab}]_q = A_\gamma h_{ab},  
\end{eqnarray}
and,
from
$
e^a_A
= \big(\frac{\partial x^a}{\partial \theta^A}\big)_\lambda
= \big(\frac{\partial x^a}{\partial \theta^A}\big)_{\tilde\lambda}
= [e^a_A]_q,
$

\begin{eqnarray}\label{sigma_q}
  [\sigma_{ab}]_q
  &=& q_{ab} [e^a_A]_q [e^b_B]_q
  \nonumber \\
  &=& [h_{ab}]_q [e^a_A]_q [e^b_B]_q
  \nonumber \\
  &=& [h_{ab}]_q e^a_A e^b_B
  \nonumber \\
  &=& A_\gamma \sigma_{ab}.
\end{eqnarray}  
The qmetric volume element
is
$
[d^{D-1}V]_q = [\sqrt{\sigma}]_q \, d^{D-2}\theta \ d{\tilde\lambda}
            = [d^{D-2} {\cal A}]_q \, d{\tilde\lambda}
$
with
$
[d^{D-2} {\cal A}]_q = [\sqrt{\sigma}]_q \, d^{D-2}\theta
$
the $(D-2)$-dim area of the element of surface
transverse to the generators
according to the qmetric,
and $d^{D-2} {\cal A} = \sqrt{\sigma} \ d^{D-2}\theta$
the area according to $g_{ab}$.
By the way,
this form of $[d^{D-1}V]_q$ gives,
from
\begin{eqnarray}
  \frac{[d^{D-1}V]_{q(g)}}{[d^{D-1}V]_{q(\eta)}} =
  \frac{[d^{D-2}{\cal A}]_{q(g)}}{[d^{D-2}{\cal A}]_{q(\eta)}},
\end{eqnarray}
an equivalent manner, if one wants, to express $\rho$,
as

\begin{eqnarray}
     \rho(P, l^a) =
  \bigg(\lim_{p \rightarrow P}
  \frac{[d^{D-2}{\cal A}]_{q(g)}(p, P)}{[d^{D-2}{\cal A}]_{q(\eta)}(p, P)}\bigg)
  _{\gamma(l^a)}. 
\end{eqnarray}  
From (\ref{sigma_q}),
\begin{eqnarray}  
  [d^{D-1}V]_q
  &=& [\sqrt{\sigma}]_q \ d^{D-2}\theta \ d{\tilde\lambda}
  \nonumber \\
  &=& A_\gamma^{\frac{D-2}{2}} \sqrt{\sigma} \ d^{D-2}\theta \ d{\tilde\lambda}
  \nonumber \\
  &=& A_\gamma^{\frac{D-2}{2}} d^{D-2} {\cal A} \ d{\tilde\lambda}.
\end{eqnarray}
Using, on the
$(D-2)$-surface, orthogonal
coordinates $z^A$ such that,
chosen any one of them, $z^{\bar A}$,
it can be put in the form $z^{\bar A} = \lambda \, \chi$,
with $\chi$ such that
$\lambda \, d\chi$ is proper distance,
we can write

\begin{eqnarray}
  \nonumber
  [d^{D-1}V]_q
  = A_\gamma^{\frac{D-2}{2}} \lambda^{D-2} \ (1 + {\cal O}(\lambda^2))
  \ (d\chi)^{D-2} d{\tilde\lambda},
\end{eqnarray}
where the ${\cal O}(\lambda^2)$ term represents the effects of
curvature and is absent in flat case.
Substituting here the expression (\ref{A_gamma}) for $A_\gamma$,
we get

\begin{eqnarray}
  [d^{D-1}V]_q
  &=& {\tilde\lambda}^{D-2} \frac{\Delta}{\Delta_S}
  \ (1 + {\cal O}(\lambda^2))
  \ (d\chi)^{D-2} d\tilde\lambda.
  \nonumber
\end{eqnarray}
Taking the limit $\lambda \rightarrow 0$ we see that this quantity,
as well as $[d^{D-2}{\cal A}]_q$,
do
not vanish, going to the values

\begin{eqnarray}
  \lim_{\lambda \rightarrow 0} \ [d^{D-1}V]_q
  = L^{D-2} \frac{1}{\Delta_L(P, l^a)} \ (d\chi)^{D-2} d\tilde\lambda,
\end{eqnarray}
and

\begin{eqnarray}
  \lim_{\lambda \rightarrow 0} \ [d^{D-2}{\cal A}]_q
  = L^{D-2} \frac{1}{\Delta_L(P, l^a)} \ (d\chi)^{D-2},
\end{eqnarray}
with
$\Delta_L(P, l^a) = \Delta({\bar p}, P)$,
where $\bar p$ is that point on the null geodesic $\gamma(l^a)$
which has $\lambda({\bar p}, P) = L$.
In the flat case,
$\Delta = 1$ identically and then $\Delta_L(P, l^a) = 1$, as we said,
and the expressions above reduces to
$
\lim_{\lambda \rightarrow 0} [d^{D-1}V]_{q(\eta)} =  L^{D-2} \
(d\chi)^{D-2} d\tilde\lambda
$
and
$
\lim_{\lambda \rightarrow 0} [d^{D-2}{\cal A}]_{q(\eta)} =  L^{D-2} \
(d\chi)^{D-2}.
$
Thus,

\begin{eqnarray}\label{Delta_null}
  \rho(P, l^a)
  &=& \frac{\lim_{\lambda \rightarrow 0} [d^{D-1}V]_{q(g)}}
  {\lim_{\lambda \rightarrow 0} [d^{D-1}V]_{q(\eta)}}
  \nonumber \\
  &=& \frac{1}{\Delta_L(P, l^a)}.
\end{eqnarray}
We obtain then, in the null case, that same form we found in the
timelike/spacelike case.
Since $l^a$ is assigned with the null geodesic at start,
we notice that,
even if the qmetric $q_{ab}$ does depend on the chosen observer (through $m^a$),
no dependence on the observer is left in $\rho$.

For timelike/spacelike geodesics,
we gave an expansion of $\Delta(p, P)$ in powers
of $l = \sqrt{\epsilon \sigma^2}$ (equation (\ref{expansion})).
For (affinely parameterized) null geodesics,
$\Delta(p, P)$ can be analogously expanded in powers
of $\lambda$ as (\cite{DeWA}; \cite{Xen, VisA, PPV})

\begin{eqnarray}
  \Delta(p, P) = 1 + \frac{1}{6} \lambda^2 R_{ab}(P) l^a l^b
                 + o(\lambda^2 R_{ab}(P) l^a l^b). 
\end{eqnarray}
For $l^a$ in a neighbourhood of $0$,
this definitely gives

\begin{eqnarray}
  \Delta_L(P, l^a) = 1 + \frac{1}{6} L^2 R_{ab}(P) l^a l^b
                     + o(L^2 R_{ab}(P) l^a l^b),
\end{eqnarray}  
and

\begin{eqnarray}
  \rho(P, l^a)
  = 1 - \frac{1}{6} L^2 R_{ab}(P) l^a l^b + o(L^2 R_{ab}(P) l^a l^b).
\end{eqnarray}  
This expression for $\rho$ is analogous
to that reported above for timelike/spacelike geodesics
(equation (\ref{rho_expansion})),
and coincides with the expression which has been found
through recourse to Euclidean sector
\cite{Pad04, Pad06, Pad08, Pad10}.

\section{Conclusions}

Starting from the quantum metric $q_{ab}$
put forward in \cite{KotE, KotF, StaA}
for timelike/spacelike intervals
from the assumption of existence
of a lower limit length
(along with some consistency conditions),
we have introduced a notion of quantum metric $q_{ab}$
for null separated events, and found an expression for it
in equation (\ref{qab_null}) (with (\ref{alpha_gamma}) and (\ref{A_gamma})).
This expression,
and the already existing expressions for timelike
and spacelike geodesics \cite{StaA},
complete the task
of providing quantum expressions
for any kind of spacetime intervals.
This quantum metric comes from a single basic request,
that of length quantization,
not from a specific quantum theory of gravity.
As such, it finds in principle wide range applicability
across any specific quantum model of gravity which foresees
quantization of length,
i.e. in practice several, if not all, models.
This means that in any such model these formulae
might be reproducible and cross-checkable.

The formulae for $q_{ab}$
for non-null intervals
hint towards
a statistical interpretation
of spacetime \cite{Pad04},
and this is exploited 
in the introduction of
a scalar function $\rho(P, v^a)$ expressing the density
of quantum states, at event $P$ in the direction $v^a$,
associated with atoms
we may think spacetime is made of \cite{Pad04, Pad06, Pad08, Pad10}.
Crucial to this, is the realization that,
according to the quantum metric $q_{ab}$
as applied to the Euclidean sector,
the cross-sectional area of an equi-geodesic surface
centered at $P$
does not vanish but goes to a finite limit,
when
the surface shrinks classically to $P$,
signalling this way (quantum) degrees
of freedom for spacetime at $P$ \cite{Pad04}.
Here, we have used
the formula for $q_{ab}$ for null separated events
to derive 
an expression for $\rho$ for $v^a$ null, 
thus remaining entirely within the Lorentz sector,
i.e. without making use of Euclideanization
(which is how $\rho$ was originally introduced).
Key to this,
has been to find out that,
analogously to what happens in the Euclidean case,
according to the null quantum metric
the cross-sectional area of a null equi-geodesic surface 
centered at $P$ does not vanish but remain finite
when the surface shrinks classically to $P$.
The formula we obtain for $\rho$
turns out to coincide
with the formula
derived through Euclideanization.
The formula for null intervals,
joined with the formulae for timelike/spacelike cases,
provide a complete account of $\rho$
based on quantum spacetime intervals.

{\it Acknowledgements.}
I am grateful to Sumanta Chakraborty and Dawood Kothawala
for remarks and discussions on the topics of the paper.


\end{document}